\documentstyle[11pt,twoside,pasp3D,epsf]{article}

\begin{document}

\title{Image Slicing with Infrared Fibers}

\author{J. E. Larkin$^1$, A. Quirrenbach$^2$, \& J. R. Graham$^3$}
\affil{$^1$Department of Physics and Astronomy, University of California,
Los Angeles, 8371 Math Sciences, Los Angeles, CA. 90095-1562}
\affil{$^2$Center for Astrophysics and Space Science, University
of California, San Diego, 9500 Gilman Drive, La Jolla, CA 92093-0111}
\affil{$^3$Department of Astronomy, University of California, Berkeley,
Berkeley, CA 94720}

\begin{abstract}

We are proposing to build a new integral field instrument (OSIRIS) for
use with the Keck Adaptive Optics System.  It will utilize a large
(1024 element) fiber optic bundle to slice the field and feed a
standard infrared spectrograph with a spectral resolution of 5000.  To
improve the fill factor of the fiber bundle, we plan on coupling it to
a matched lenslet array.  The spectrograph will have three plate
scales of 0\farcs05, 0\farcs10 and 0\farcs20 per pixel, and full broad
band spectra will fit on a single 2048$\times$2048 infrared array.
Other innovations include using some of the fibers for pseudo-slits
and sky measurements and non-uniform spacing for the "linear" feed to
the spectrograph.

\end{abstract}

\section{Design Goals}

Among the most important design goals for the OSIRIS instrument was to
have a sampling close to the diffraction limit of the Keck Telescope
(0\farcs02 to 0\farcs07 in the near-IR).  The only ways of achieving
this is to either have very small physical samping (0\farcs05 =
36.4~$\mu$m at F/15) or a large focal ratio (0\farcs05 = 1~mm at
F/412).  The first option requires new technologies and one must worry
about diffraction effects deteriorating image quality.  The second
option make the spectrograph very large and potentially costly.  As
you'll see, we've opted for a compromise of moderate focal ratio and
sampling size.

Many of the potential science targets are on the order of one
arcsecond across and one of our design goals was to encompass all or
most of a target in a single exposure.  With samplings on the order of
0\farcs05, this requires at least a 20$\times$20 grid of spatial
samplings or at least 400 individual spectra.  To avoid spectral
overlap, for most geometries this requires at least a 1024 square
array.

The instrument also needs to have a spectral resolution of at least
4000.  This requirement is set both by the spectral sampling needed on
many of the science targets ($\sim$ 75 km/s) and also on our desire to
suppress OH emission lines.  Between 1 and 2.2 $\mu$m the dominant
source of background is the forest of atmospheric OH lines
(e.g. Herbst, 1994). The continuum emission actually reaches a minimum near
1.6 microns where scattered light and thermal emission cross. Because
the OH lines are narrow and discrete, we can separate them by going to
high spectral resolution. At a resolution of 4000 approximately 90\%
of our pixels will be free of bright emission lines and will thus have
greatly reduced backgrounds.

A conflicting requirement is broad band spectral coverage (z, J, H and K).
This is an efficiency requirement so that each setting has $\Delta\lambda /
\lambda \sim 0.2$.  Coupled with the high spectral resolution, this requires
approximately 2000 spectral pixels.

Finally, the spectrograph needs to work throughout the near infrared from
1 to 2.5 microns are therefore requires full cryogenic performance (77 K).
This requirements places many difficulties with the design including
the need for cryogenic optical couplings, cryogenic motors and mechanisms,
temperature compensation in the optical alignment and restrictions on
the optical components that are available.

\section{Image Slicing Technique}

With the requirement of diffraction limited sampling mentioned above,
we found that we had three viable image slicing techniques.  The first
would be to use micromirrors (MEMs) to carve up the field.  This is a
promising avenue, but we have not seen any concrete designs for
coupling a micromirror to a spectrograph.  One of the largest problems
with these devices is that each mirror tips about an individual axis
so the optical path lengths from different elements are different.
This makes it difficult to collimate the light. Micromirrors also
require complex electronics and software to control. Other concerns
such as fill factor, repeatibility, cryogenic operation and technical
maturity also made us decide against micromirrors.

A second slicing solution would be the use of a lenslet array in a
Tiger style instrument (Bacon et al. 1995). That is use the pupil images
formed by a lenslet array placed at the focus to feed a spectrograph.
The lenslets are rotated so that their spectra don't overlap.  This
is a good option if the desire is for large spatial coverage with modest
spectral coverage.  The difficulty arises in that spectra from the
bottom of the array must be prevented from overlapping with spectra
from the top or middle of the array.  One solution would be to make
a fairly long and narrow lenslet array (10$\times$50 for example) but
that would not meet our goal of 1'' spatial coverage in both axes.

The third slicing option we considered is a fiber optic bundle.  These
have now matured into a somewhat standard commercial product in which
fibers can be arranged in complex geometric patterns at the input, and
then reorganized into a linear (or pseudo-linear) pattern for feeding
a spectrograph.  They offer the advantages of a fine sampling scale (100 $\mu$m), no spectral overlap between fibers, and the ability to dedicate some
fibers to simultaneous measurements of the sky.  The primary disadvantage
is modest fill factors ($\sim$40\%).  To compensate for this we have
combined options two and three and designed around a fiber bundle and matched
lenslet array placed at the AO focus. The lenslet is used to achieve
a much higher fill factor, while the fibers gives increased sampling
flexiblity and a more optimized arrangement into the spectrograph.

Fiberoptic manufacturers have begun creating elaborate fiber bundles
for high speed switching applications, and we have identified one
company, Fiberguide Industries, that can produce almost arbitrary
fiber patterns.  The basic process is to define a hole pattern in a
computer and use it to direct a laser to cut a precision set of holes
in a thin metal plate.  The hole spacing can be as small as 150
microns and positional accuracies are better than 4 microns.  The
fibers are then individually polished and inserted into the holes.
The output end is done similarly with essentially any mapping of input
to output hole patterns.  Fiberguide Industries has an infrared
transmissive fiber (Anyhroguide G) that operates from 0.4 to 2.4
microns and has a temperature range from -190\( ^{\circ } \)C to
350\( ^{\circ } \)C; ideal for our application.  We
anticipate a fully cooled optical assembly to just above LN2
temperatures.

\begin{figure}
\plotfiddle{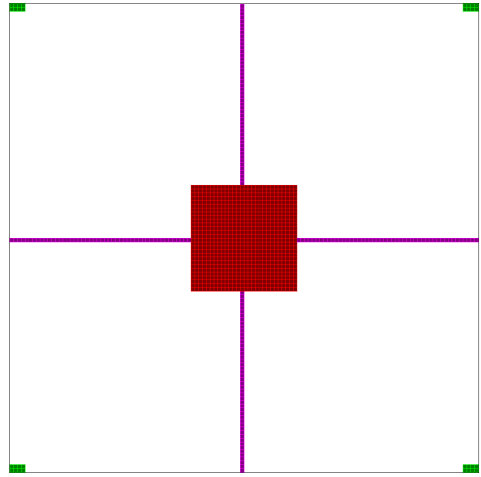}{2.4in}{0}{40}{40}{-120}{-70}
\caption{Input fiber pattern showing central 28$\times$28 bundle, 132$\times$1
horizontal and vertical "slits" and the four sky patches at the
extreme corners. \label{fig:input}}
\end{figure}

Based on certain optical limitations it is difficult to use more than
one fiber per two spatial pixels, even with underfilling of each
pixel.  We are therefore proposing a 1024 fiber bundle with a fairly
elaborate input and output arrangement.  The primary input component
will be a 28$\times$28 fiber rectangular bundle (784 fibers total).
Extending from this bundle in the four cardinal directions are 52
fibers yielding pseudo-slits of total length $52+28+52= 132$ fibers.
These "slits" can be aligned along an extended object's major and
minor axis to extend the spatial coverage of the instrument.  The
final 32 fibers are placed in four bundles of 4$\times$2 fibers each
at the extreme corners to provide simultaneous sky coverage for many
compact objects.  The arrangement is shown in figure~\ref{fig:input}.

At the output the fibers are staggered to allow a horizontal
spacing of 100 microns while maintaining the minimum 150 micron
fiber-fiber spacing.  Since an arbitrary fiber spacing is possible
with bundle technology, we plan on an output pattern that is not
exactly evenly spaced and that has a gentle curve from one end to the
other.  This allows us to remove much of the spectral curvature and
distortion of the optics that is inherent in slit spectrographs.  So
it is hoped that the spectra on the detector will have a very uniform
spacing and spectral features will lie roughly along rows of the
device.

\subsection{Lenslets}

A fundamental problem with fiber bundles is the low fill factor due to
the round shape of the fibers and the thick claddings and jackets.  We
are proposing to remove this problem by placing a matched lenslet
array in front of the fiber bundle. The fibers are then at the pupil
images of the lenslets making the instrument a pupil spectrograph.
Adaptive Optics Associates can manufacture custom lenslets with the
desired 150 micron spacing with better accuracy than the fiber bundle.
Rectangular lenslets have $>98\%$ fill factor and should have high
coupling efficiency.  For the infrared, AOA recommends their
sapphire substrates with the lenslets formed from infrared
transmissive epoxy.  AR coatings will be applied to both faces.

A potential problem is the alignment of the lenslets with the fiber
bundle and maintaining this alignment when cold.  We hope to alleviate
this problem by rigidly mounting the lenslets to the face of the fiber
bundle.  The lenslets can be fabricated such that their focal plane is
at their back surface.  By fastening this to the fibers, alignment
should remain good during cooldown.  We will investigate several
fastening strategies including ir-epoxy, and mechanical clamping.

For most of the plate scales we recommend, the alignment requirement
is relaxed a bit because the spot size generated by each lenslet is
significantly smaller than the fiber core itself.  So we underfill
each fiber and small misalignments don't create light losses.  The
spot size for each lenslet is essentially the pupil diameter formed
by the lenslet since each lenslet is in the focal plane.  The pupil
size is given by:

$$ pupil\,\, size(microns) = {lenslet\,\, focal\,\, length(microns)
\over input\,\, focal\,\, ratio} $$

As shown below, the input focal ratio for the primary fiber spacing of
0\farcs05 per fiber is F/61.  To achieve a pupil size under the 50
micron core size, we need a lenslet focal length less than 3.05 mm.
For the coarser plate scales of 0\farcs1 and 0\farcs2 per fiber, the
pupil would be too big, however, resulting in a loss of light.  At the
same time too small of a focal length will result in a very fast focal
ratio at the detector.  We are therefore recommending a focal length
of 1.5 mm, which matches the 0\farcs1 pupil with the fiber core.  It
does mean that there will be a 75\% loss of light for the coarsest
scale of 0\farcs2 per fiber.  As will be shown, a focal length of 1.5
mm for the lenslets requires a focal ratio at the detector of F/3.6
which is difficult but doable.

\section{Spectrograph Design}

\subsection{Front-End Reimaging Optics}

The angular scale that corresponds to each fiber is determined by the
F/\# of the beam as it reaches the lenslet array.  A simple
relationship can be obtained:

$$ { \Delta\theta('') \over \Delta x (\mu m)} = { 0.206 \over D(m) F/\#} $$

\noindent
where D(m) is the diameter of the telescope in meters, F/\# is the
focal ratio at the lenslets, and $\Delta\theta / \Delta x$ is the
angular scale in arcseconds per micron.  So for a 150 micron lenslet
separation, a 10 meter telescope and a focal ratio of 15, the angular
scale per lenslet is 0.206 arcseconds.  This is a good sampling for a
non-AO mode, but with a diffraction limit at 1 micron of under 0.025
arcseconds, it is necessary to change the plate scale before the
lenslet array.  It is also difficult to form a good cold pupil or
baffle system after the fiber bundle, so the primary baffling is
performed in the front-end optical elements.

\begin{table}[hbtp]
\centering
{\small \begin{tabular}{ccccc}
\multicolumn{5}{c}{\bf TABLE \ref{t:scales}} \\
\multicolumn{5}{c}{\bf Plate Scales} \\
\hline
& Focal Ratio & Angular Scale & & Length \\
Magnification & at Lenslets & per Fiber & Central Field & of ``slits'' \\
\hline
1:1 & F/15 & 0\farcs206 & 5\farcs76 & 27\farcs2 \\
1:2 & F/31 & 0\farcs100 & 2\farcs8 & 13\farcs2 \\
1:4 & F/62 & 0\farcs050 & 1\farcs4 & 6\farcs6 \\
1:10 & F/155 & 0\farcs020 & 0\farcs56 & 2\farcs64 \\
\hline
\end{tabular}}
\label{t:scales}
\end{table}

Table \ref{t:scales} gives the four possible plate scales and their
corresponding focal ratios.  The fields of view of the central
28$\times$28 fiber patch and the lengths of the "slits" formed by long
1$\times$132 fiber stripes is also given in arcseconds.  For full
sampling of the diffraction limit, we would need a scale per fiber of
$\sim$0\farcs01 but then the field of view becomes very small, and we
believe that for that scale, the other AO instrument NIRC2 is more
appropriate in most scientific applications.  But a field of 0\farcs56
is actually not too small to consider for certain applications like
faint field galaxies where the diameters can be under 0\farcs5.  But
the scale that we recommend as the primary choice is a 1:4
magnification that yields 0\farcs05 fibers and a 1\farcs4 central
field of view.  This is ideal for most of the scientific programs and
although it slightly undersamples the diffraction limit at all of the
selected wavelengths, it does maximize the energy per fiber for point
sources and allows for very high resolution on extended sources. The
"slit" lengths of 6\farcs6 are also quite reasonable for investigating
the major and minor axes of extended sources such as AGN and the
nuclei of Ultraluminous Infrared Galaxies.

There will, however, be occasions where a larger field of view will be
desirable so coarser samplings should be considered.  Since the front
end will consist of a pair of infrared doublets that are of very low
cost compared to the overall instrument, we are recommending that a
simple turret be used to rotate in different reimaging optics to
select different plate scales.  We recommend three selectable scales
for the instrument of 0.2, 0.1 and 0.05 arcseconds per fiber in a
three position turret.  Good repeatibility of this mechanism is
important but not critical since any motion of the lens assemblies
would simply change the field of the fiber bundle by a small
amount. So no spectral shifts or loss of light would occur due to
position errors.  The turret also provides the future option that
additional scales can be created, and three selected for each run.
This would give a powerful non-AO capability to the instrument as
well, although there can be significant light loss at the fiber
coupling with the coarsest plate scales.

\subsection{Grating}

\begin{figure}
\plotfiddle{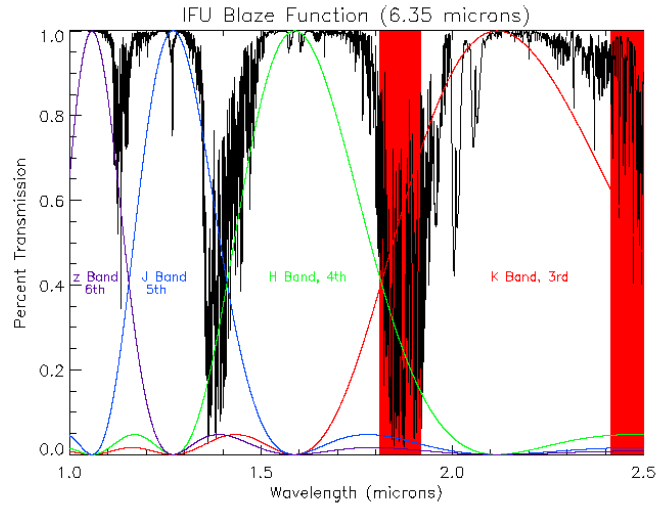}{2.6in}{0}{70}{70}{-230}{-190}
\caption{The blaze function for a grating with a blaze wavelength of
6.35 microns. Note how well each order fits within the atmospheric
transmission. \label{fig:blaze}}
\end{figure}

For maximum reproducibility, we have selected to use a single
non-rotating grating.  But we still desire high efficiency across each
infrared spectral band, and no order overlap within each broad
band. This is achievable by blazing the grating to a blaze wavelength
of 6.35 microns.  The third order then peaks at 2.12 microns, fourth
at 1.59 microns, fifth at 1.27 microns and sixth at 1.06 microns.
These are ideally situated in the K, H, J and z bands, respectively.
A blaze function for such a grating is shown in figure \ref{fig:blaze}.

\subsection{Filters}

Since an entire broad band fits onto the detector at once, narrow band
filters are not needed.  So the only filters required are broad band K
(2.0-2.4 microns), H (1.5-1.85 microns), J (1.15-1.35 microns) and z
(1.00-1.12 microns).  We plan on placing these in a filter wheel
just in front of the grating, essentially at the pupil plane.  This
gives uniform wavelength blocking across the field.  Ray tracings for
standard one inch filters have demonstrated that flatness requirements
are quite modest since filters begin as highly polished substrates.
Most surface deviations are then due to warping which are matched
between the front and back surface.

\subsection{Detector}

Essentially every advantage of the integral field unit is tied to the
use of a 2048$\times$2048 infrared device.  Good OH-suppression requires a
spectral resolution of 4000 or higher. It also requires a highly
reproducible spectral format to assist with software extraction of
uncontaminated pixels.  Together these requirements mean full band
coverage without a moving the grating.  For a resolution of 5000, this
requires roughly 2000 pixels in each spectral channel.  If a 1024
array were used, then either the resolution would need to be halved
with considerable loss of scientific capability and essentially all
OH-suppression, or the grating would need to be scanned across the
array which would decrease observing efficiency and degrade spectral
extractions.

The spatial coverage is also dependent on the number of fibers in the
bundle.  We've found it difficult to use more than one fiber for every
two columns of the array without significant spectral overlap.  If a
1024 array were used, then this would correspond to only 512 fibers
arranged in a central 20$\times$20 fiber bundle.  With 0\farcs05 fiber
sampling, the field would be one arcsecond on a side.  This is
marginal for many of the science goals.  A 2048 array would allow
1024 fibers arranged in the 28$\times$28 fiber bundle with 240 fibers for the
extended linear "slits" and sky patches described above.  This is a
significant improvement in field size and relaxes many of the pointing
accuracies required for targets such as the Kuiper belt objects.

Rockwell has recently begun fabricating multiplexers for a
new Hawaii 2 infrared array that has a 2048$\times$2048 format.
Quantum efficiencies are expected to be greater than 80\% for the 1-2.5
micron version, with a read noise of $\sim$2.5 electrons.  The detector
pitch will be 18 microns yielding a total array physical size of 37
millimeters.  The multiplexer can support either 4 or 32 readout
channels with pixel clocking rates of roughly a megahertz.  Since all
of science goals make use of the OH-suppression, it is anticipated
that long exposure times will be the rule, so it is not necessary to
support a fast readout mode.  We therefore believe that a one hertz
maximum frame rate is more than sufficient allowing us to use only 4
readout channels.  This greatly reduces the cost of the electronics.

\section{Science Drivers}

The OSIRIS instrument combines four powerful capabilities: high
spatial resolution with the Keck adaptive optics system, medium
resolution (R$\sim$5000) infrared spectroscopy, integral field
capability, and OH-suppression.  Taken together these capabilities
allow us to dig much deeper into many pressing astrophysical problems.
Below we have identified several scientific programs which should
greatly benefit from this instrument.

\subsection{Extragalactic Science}

Faint field galaxies like those observed in deep imaging surveys are
very compact, often with radii well below 1''.  Integral field
observations of these objects can determine internal dynamics, star
formation rates, redshifts, and Hubble Type simultaneously.  These
observations are crucial in determining the merger rates and
evolutionary processes in the first generations of galaxies.  Although
the surface brightness per fiber will be quite low, the backgrounds
are also down by a factor of 400 compared to a square arcsecond so
many of these objects can be observed in reasonable timescales (hours).

\begin{figure}
\plotfiddle{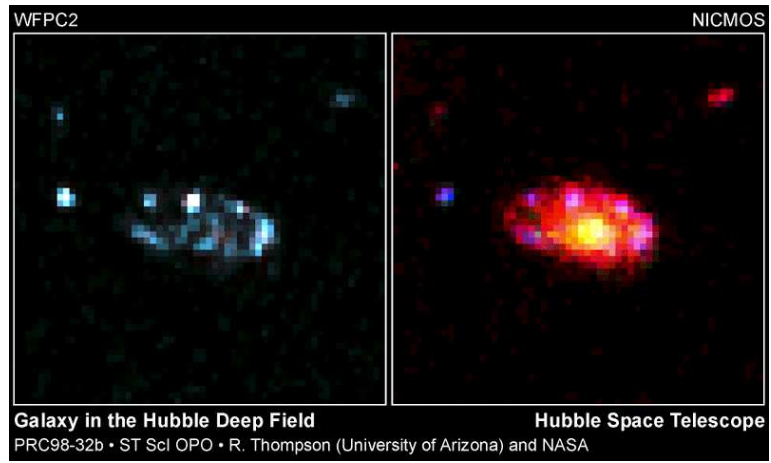}{2.4in}{0}{65}{65}{-195}{-170}
\caption{Faint field galaxies become very small at high redshift and
can have core diameters well below 1''. As this pair of images from Roger
Thompson (Univ. of Arizona) shows, the infrared and optical morphologies
are sometimes quite different.\label{fig:faint}}
\end{figure}

Another related group of objects are the host galaxies of quasars.  Only
with high spatial resolution can the faint disks of galaxies be seen in
the glare of quasars.  Recent observations suggest that mergers and
galaxy interactions may play a key role in the feeding of material in
the central supermassive blackhole.  An integral field instrument
can study not only the emission features of the quasar itself, but
also measure the morphology, and spectrum of the galaxy and determine
star formation rates, masses, metallicities, etc...

\begin{figure}[htb]
\plotfiddle{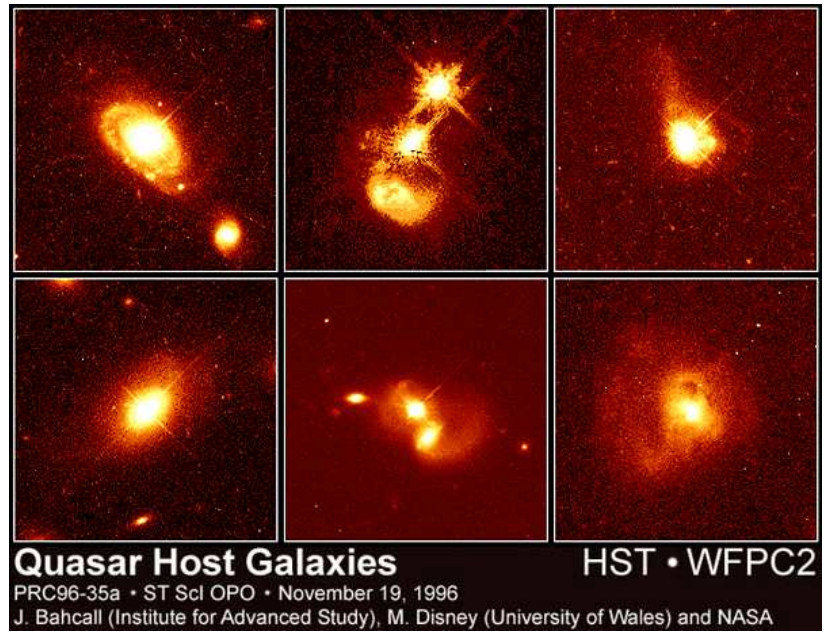}{3.4in}{0}{60}{60}{-185}{-115}
\caption{As these recent Hubble Space Telescope images show, the host
galaxies of quasars are often highly disturbed and very compact. Only
with integral field spectroscopy on subarcsecond scales can these
galaxies be studied in detail.
\label{fig:qso}}
\end{figure}

Active Galaxies are also an ideal target for the integral field
spectrograph.  At the resolution of the proposed instrument
(0\farcs05) the narrow line regions of some nearby Seyfert Galaxies
should be resolvable.  This would allow us to directly test the
standard model in which AGN have a supermassive blackhole, an
accretion disk and two populations of clouds (narrow line and broad
line) which orbit at different distances from the center. OSIRIS
should be able to spatially and spectrally separate out the narrow
line clouds from the more centralized broad line region.  OSIRIS
should also be able to study nuclear star formation in AGN and
spectrally image the interaction between the central engine and
surrounding regions.

\subsection{Planetary Science}

Although the planets are vastly closer than the extragalactic targets,
many solar system targets do have interesting structures on subarcsecond
scales.  Obvious targets include the moons of the Jovian planets and
asteroids.  The largest moons, such as Titan are roughly 1'' across, and
there is intense interest in their surface chemisty. Titan has a dense
methane cloud layer which can best be penetrated in the infrared, and
appears to have a surface covered in hydrocarbons.  With OSIRIS, one could
measure surface chemistry at over 500 independent locations within a
few hours.  The moons IO and Ganymede are also of great interest.

Kuiper belt objects present a different opportunity for OSIRIS. Although
we don't anticipate resolving these tiny objects, there is still significant
gain to be had by going to high spatial resolution. Chief of these is
a great reduction in the background. The integral field nature of OSIRIS
also reduces the need for accurate tracking and positioning since there
are effectively no slit losses.

\subsection{Galactic Objects}

Within our Galaxy, there are countless varieties of objects, many of
which have structural and spectral properties ideally suited for
OSIRIS. Perhaps chief among these are faint stellar objects and brown
dwarfs.  For free floating objects, the reduced background of the high
spatial resolution would allow for more sensitive measurements.  For
faint objects near brighter companions, the pseudo-coronagraphic
nature of an integral field instrument (and the capability to put in a
real coronagraphic set of stops) gives great advantage in determining
the spectral properties.  OSIRIS not only offers the ability to study
such objects but also to find them in the first place.  By imaging the
central 1'' to 2'' around stars, one could use a spectral filter to
search for objects that are spectrally different from the primary.
Such a filter could be formed from a weighted summed set of spectral
channels where the brown dwarf or low mass companion would be expected
to be significantly different from the primary, such as in the methane
or water bands.

The Galactic center is a unique environment where massive stars are born
in compact clusters, and where orbital motion is directly observable around
the central blackhole.  It is also highly obscured with approximately
30 magnitudes of visual extinction.  Infrared spectra encompassing the
central cluster ($<$1'' in diameter) could determine the 3-dimensional
kinematics as well as the spectral types of this unique grouping of stars.

\section{Instrument Summary}

The OSIRIS instrument is an integral field infrared spectrograph
designed for the Keck Adaptive Optics System.  It will utilizes the
latest infrared detectors and state-of-the-art fiber bundle
technologies to achieve high spatial resolution, integral field
imaging, very faint object sensitivity, moderate resolution
spectroscopy (R$\sim$5000) and a limited long slit capability.  All of this
is achieved with an innovative yet straightforward and easy to use
instrument.

Because of the great amount of information in each frame, only very
limited observing modes are required.  The large array means that a
full broad band spectrum is contained in each exposure. By proper
grating selection this is further simplified in that a single fixed
grating can be used for all bands (z, J, H and K).  The band is easily
selected by changing the order sorting broad band filter.  Thus,
within the spectrograph portion of the instrument, the only moving
part is a simple filter wheel with five positions corresponding to z
band, J band, H band, K band and a closed position.  In each position,
a complete spectral image is obtained with 2048 spectral channels and
1024 spatial channels.

Within each exposure is an R$\sim$5000 spectrum of a 28$\times$28
pixel patch of the sky, plus vertical and horizontal slit coverage of
1$\times$132 pixels.  At this resolution, OH emission lines that
dominate the background in the J and H windows are well separated. By
coadding only those spectral channels without OH-emission lines (about
90\% are clean) a very deep infrared image can be extracted for the
28$\times$28 pixel field (1.4''$\times$1.4'' in the primary adaptive
optics mode).  But in addition to this image, full spectral coverage
is of course present.

The OSIRIS instrument is not only powerful scientifically but also
quite modest in cost and complexity.  It draws upon much of the
experience of previous Keck instruments especially NIRSPEC (McLean et
al, 1998).  Due to a simpler detector readout, it actually has simpler
electronics requirements than the NIRSPEC instrument allowing us to
essentially clone most of the vital electronics.  Because of the
expected low backgrounds and long exposures, the detector can be
clocked out relatively slowly, about 1 hertz per frame, meaning that
we only need to use a four channel readout scheme.  We also anticipate
only two or perhaps three cryogenic mechanisms of fairly simple
design.

\acknowledgments

The OSIRIS design study is supported by a grant from the California
Association for Research in Astronomy which owns and operates the Keck
Observatory.


\begin{references}
\reference Bacon, R., et al., 1995 \aaps, 113, 347
\reference Herbst, T. M., 1994 \pasp, 106, 1298
\reference McLean, I. S., et al., 1998, SPIE Proc., 3354, 566
\end{references}
\end{document}